\documentclass[conference]{IEEEtran}% \usepackage{cite}

\usepackage{cite}
\usepackage{amsmath,amssymb,amsfonts}
\usepackage{algorithmic}
\usepackage{graphicx}
\usepackage{textcomp}
\usepackage{xcolor}
\usepackage{multirow}%
\usepackage{mathrsfs}%
\usepackage{manyfoot}%
\usepackage{booktabs}%

\usepackage{hyperref}

\begin{document}

\title{Survival of the Cheapest: Cost-Aware Hardware Adaptation for Adversarial Robustness
\thanks{Financial support has been provided in part by the Knut and Alice Wallenberg Foundation grant number 2019.0352 and by the eSSENCE Programme under the Swedish Government's Strategic Research Initiative.}
}

% \author{Anonymous Authors}

\author{\IEEEauthorblockN{1\textsuperscript{st} Charles Meyers}
\IEEEauthorblockA{\textit{Responsible AI Practice } \\
\textit{Northeastern University}\\
Boston, USA \\
0000-0002-1277-9811 \\
cmeyers@cs.umu.se}
\and
\IEEEauthorblockN{2\textsuperscript{nd} Mohammad Reza Saleh Sedghpour}
\IEEEauthorblockA{
\textit{DoiT}\\
Lund, Sweden \\
0000-0002-0751-9695}
\and
\IEEEauthorblockN{3\textsuperscript{rd} Erik Elmroth}
\IEEEauthorblockA{\textit{dept. Computer Science} \\
\textit{Ume\aa \space University}\\
Ume\aa, Sweden \\
0000-0002-2633-6798
}
\and
\IEEEauthorblockN{4\textsuperscript{th} Tommy L\"{o}fstedt}
\IEEEauthorblockA{\textit{dept. Computer Science} \\
\textit{Ume\aa \space University}\\
Ume\aa, Sweden \\
0000-0001-7119-7646}
}

\maketitle

\begin{abstract}
Deploying adversarially robust machine learning systems requires continuous trade-offs between robustness, cost, and latency.
We present an autonomic decision-support framework providing a quantitative foundation for adaptive hardware selection and hyper-parameter tuning in cloud-native deep learning.
The framework applies accelerated failure time (AFT) models to quantify the effect of hardware choice, batch size, epochs, and validation accuracy on model survival time.
This framework can be naturally integrated into an autonomic control loop (monitor--analyse--plan--execute, MAPE-K), where system metrics such as cost, robustness, and latency are continuously evaluated and used to adapt model configurations and hardware selection.
Experiments across three GPU architectures confirm the framework is both sound and cost-effective: the Nvidia L4 yields a 20\% increase in adversarial survival time while costing 75\% less than the V100, demonstrating that expensive hardware does not necessarily improve robustness. 
The analysis further reveals that model inference latency is a stronger predictor of adversarial robustness than training time or hardware configuration.
\end{abstract}


\section{Introduction}
Cloud-native deployments of machine learning with deep neural networks are increasingly common in safety-critical classification tasks, with applications ranging from medical imaging~\cite{ai_medical_imaging} to aviation~\cite{ai_aviation} and from security~\cite{ai_security,ai_luggage,ai_prison} to self-driving cars~\cite{ai_automotive}.
Statistical learning theory~\cite{vcdimension,shalev2014understanding} provides no guarantees about the generalisation performance of deep neural networks due to the massive number of tunable parameters.
To overcome this, neural networks need large amounts of data~\cite{desislavov2021compute,bailly2022effects} to train ever-larger models~\cite{desislavov2021compute}, which has yielded increasingly marginal gains on validation accuracy~\cite{sun2017revisiting}.
It is also clear that reaching safety-critical standards using validation accuracy would require an infeasibly large data set~\cite{meyers}.

Modern neural networks are massive, from AlexNet's 60M parameters~\cite{alom2018history} to Mamba's 8~billion~\cite{mamba}, and now rank among the largest consumers of data-centre power~\cite{msft_water}. 
At these scales, meeting even the weakest safety-critical threshold ($[10^{-12}, 10^{-15}]$ per second failure rate~\cite{IEC61508,iso26262,aviation_software,safetyframework}) would require billions of validation samples per model change --- computationally infeasible~\cite{meyers}.

In this work, we propose an autonomic decision-support framework based on accelerated failure-time (AFT) methods to predict model performance across hardware configurations and estimate deployment cost.

\subsection{Classifiers}

We consider ML classifiers, $K(x;\theta)$, with parameters $\theta$, for a mini-batch, $x$, of size $N$ with corresponding true labels $y$, and predictions $\hat{y} = K(x;\theta)$. 
A loss function, $L(y,\hat{y})$, quantifies the prediction error.

Due to model complexity, large-scale ML training typically uses mini-batch stochastic gradient descent (SGD), which updates parameters using mini-batch gradients over multiple epochs,
\begin{equation}
    \theta^{(i+1)} = \theta^{(i)} - \eta^{(i)} \nabla_{\theta^{(i)}} L\big(y, K(x, \theta^{(i)})\big),
\end{equation}
where $\eta^{(i)}$ is the learning rate and the gradient is computed over the mini-batch.

Hardware differences (\textit{e.g.,} VRAM) influence optimal configurations, especially when considering performance and cost. 
Training and inference may also run on different hardware to optimise efficiency, as some devices are specialised for training (\textit{e.g.,} V100, P100) and others for inference (\textit{e.g.,} L4).
An effective learning rate ($\eta$) balances fast convergence and accuracy, but in cloud environments, hardware variability requires empirical tuning.
Since compute is billed per hour, optimising training time is also cost-critical.

\subsection{Adversarial Attacks}
\label{attacks}

In this work, we focus on \textit{evasion attacks}, where the attacker perturbs the input at inference time to induce a misclassification. 
Formally, \textit{adversarial success} or \textit{failure} is one in which
\begin{equation} \label{eq:adv_success}
    K(x; \theta ) = \hat{y} \neq \hat{y}_a = K(x + \varepsilon; \theta),
\end{equation}
where $\varepsilon$ is added adversarial noise,  bounded as $0 < \|\varepsilon_i\| \leq \varepsilon^*$.
Additionally, one can measure the accuracy of the model when tasked with these adversarial samples, $x + \varepsilon$, giving a metric called \textit{adversarial accuracy}, which is Eq.~\ref{eq:acc} calculated on the perturbed samples, \textit{i.e.}, on $x + \varepsilon$. 
These perturbations may represent malicious evasion attacks or naturally occurring input deviations, but the objective is not to model a specific attacker.
Instead, these attacks provide an efficient mechanism for measuring how quickly a model fails as input conditions deviate from the test distribution.

One of many possible attacks is called the \textit{fast gradient method}~\cite{fgm}. 
It works by applying noise to a set of samples, $x$, to generate adversarial examples, $x_a$, such that,
\begin{equation}
    x_a = x + \eta \cdot \mathrm{sign}\Big(\nabla_x L\big(y, K(x, \theta)\big)\Big).
\label{eq:fgm}
\end{equation}
Although presented here as a representative example, this attack is used throughout this work. 
The methods and analysis, however, generalize to any adversarial attack when there is a multiplicative relationship between explanatory covariates (\textit{e.g.,} magnitude of the noise vector) and probability of failure~\cite{meyers_aft,aft_models,kleinbaum1996survival}.

\subsection{Related Work}

Selecting hyper-parameters for adversarially robust models is considerably more challenging than for conventional training because benign accuracy and adversarial robustness are often competing objectives~\cite{carlini_towards_2017,meyers,dohmatob_generalized_2019}. 
Duesterwald et al.~\cite{duesterwald2019} demonstrated that adversarial training is highly sensitive to hyper-parameter selection and that no single configuration performs well across datasets or robustness objectives. 
In particular, robustness verification is shown to be NP-hard~\cite{carlini_towards_2017}.

Our work differs from previous robustness-oriented HPO studies in two important respects. 
First, rather than optimising solely for benign and adversarial accuracy, we additionally optimise for training efficiency. 
This is particularly important in practical deployments where the optimal batch size is constrained by the available GPU memory, numerical precision, and model size. 
Second, we focus on identifying hardware-aware training configurations that minimise wall-clock training time while maintaining competitive benign and adversarial performance.

This paper is complementary to the TRASHFIRE framework~\cite{meyers_aft}, which proposes survival models (detailed in the next section) for evaluating the cost-efficacy of different models and attacks.

Whereas TRASHFIRE justifies and defines training-time heuristics that generalise across experiments, this work formulates hyper-parameter selection as a multi-objective Bayesian optimisation problem across hardware configurations while measuring real-world cost in terms of time, power, and money.
Specifically, we employ the Tree-structured Parzen Estimator because it has been shown to reach competitive Pareto-optimal solutions in hundreds rather than thousands of trials compared with evolutionary multi-objective optimisers such as CMA-ES and NSGA-II~\cite{ozaki2020multiobjective,optuna,tpe_params}.

\subsection{Contributions}

To tackle the problems of scaling validation to safety critical levels, we present a methodology (Section~\ref{cost}) that:
\begin{itemize}
    \item Provides a quantitative decision-support component for self-adaptive systems, showing how survival models allow model builders to derive cost and robustness estimates across the feasible hyper-parameter space.
    \item Demonstrates a scalable and effective method to train a model while simultaneously estimating the effect of various hyper-parameters.
    \item Measures the power and monetary cost of deploying a model across different hardware architectures to model the trade-offs between deployment hardware and robustness.
    \item Provide a fast rejection method for eliminating bad candidate models (Equation~\ref{eq:aft_time}).
    \item Demonstrates that ``inference-only'' GPUs are both cheap and effective for large numbers of datasets and many classes of models. 
\end{itemize}
This paper focuses on the Monitor, Analyse, and Plan phases of MAPE-K\@; live Execute-phase integration with a running autonomic system will be addressed in future work.

\section{Survival Analysis for Cost-Aware Robustness Estimation}
\label{aft}

AFT models are statistical models used to analyse multivariate effects on the observed failure rate to predict the time-to-failure across a wide variety of circumstances~\cite{aft_models,kleinbaum1996survival}, mapping the relationship between model tuning parameters and performance to provide an analytical foundation for runtime hardware selection in self-adaptive systems.
The section below first discusses the drawbacks of using accuracy as a measure of failure probability, then motivates a move towards time-domain estimates of failure rate, and then outlines methods for comparing these time-domain models to find one that fits best.

\subsection{Accuracy}
\label{acc}

Accuracy measures the frequentist probability of a failure. 
Throughout, we use the terminology \textit{benign} accuracy to refer to the performance on a data set using unperturbed data and \textit{adversarial} to refer to the performance in the presence of additive adversarial noise that is intended to confuse the model. 
The accuracy, $\lambda$, is defined as
\begin{equation}
    \lambda:= \mathrm{Accuracy} := 1 - \frac{\mathrm{False~Classifications}}{\mathrm{Total~Classifications}},
    \label{eq:acc}
\end{equation}
which is generally assumed to indicate the rate of successes in real-world data sampled from the same distribution as the training data~\cite{tan2021critical}. 
However, it ignores run-time cost~\cite{desislavov2021compute,bailly2022effects} and adversarial noise~\cite{croce_reliable_2020}. 
Furthermore, adversarial counter-examples consistently expose the benign accuracy as optimistic~\cite{carlini_towards_2017,adversarialpatch,pixelattack,biggio_poisoning_2013,meyers} in the presence of small amounts of noise~\cite{pixelattack}.

\subsection{Failure Rate}
\label{failure_rate}

The failure rate refers to the percentage or proportion of examples that cause the targeted ML model to misclassify or produce incorrect outputs~\cite{meyers}. 
To encompass the cost of a particular model or attack, the proposed methodology considers failures to be a function over some time interval (\textit{e.g.,} attack generation time) with parameters, $\theta$, so that the failure rate is the average time until a failure in a time interval around time, $t$, such that

\[
    h_{\theta}(t) := \lim_{\Delta t \rightarrow 0} \frac{p\big(\textrm{False~Classification~in~} (t, t + \Delta t] \mid \theta\big)}{\Delta t},
\]
where
$p\big(\textrm{False~Classification~in~} (t, t + \Delta t] \mid \theta\big)$
is the probability of a false classification within a small (half-open) interval around time $t$, given a particular set of model parameters, $\theta$, the $\Delta t$ is a time interval length, and $t$ is a point in time.

\subsection{AFT Models}
\label{survival_time}

AFT models are widely used in industrial, medical, and risk-mitigation contexts~\cite{kleinbaum1996survival,aft_models} to model covariate effects on expected time-to-failure; here we apply them to ML by inducing adversarial failures to measure generalisation performance under varying hardware configurations and model/attack hyper-parameters.

We model the \textit{survival time}, $S_{\theta}(t)$, as a function of time, $t$, and some set of model parameters, $\theta$, such that,
\[
    S_{\theta}(t) = p(T>t \mid \theta) = \exp\left(-\int_0^t h_{\theta}(u) \, du\right)
\]
where $p(T>t \mid \theta)$ is the probability that a model has not failed by time $t$ (``survives'' beyond time $t$). 
The expected survival time is
\begin{equation}
	\mathbb{E}_{S_\theta}[T] = \int_0^{t^*} S_\theta(u) \,du,
    \label{eq:expectation}
\end{equation}
where $t^*$ is the latest time observed in the survival data. 
However, modelling $S_{\theta}(t)$ requires a choice of modelling function for $S_{\theta}$, a function of $u$. 
The Log-Logistic, Log-Normal, and Weibull functions are widely used alternative modelling functions~\cite{kleinbaum1996survival,meyers_aft}. 
For each trial, one can measure the attack generation time to define the time interval and the adversarial accuracy to estimate the number of failures and successes in that time interval.

\subsubsection{Survival Time}
\label{accelerated}
By using adversarial samples (see Equation~\ref{eq:adv_success}), 
The likelihood of those failures depends on the amount of adversarial noise, $\varepsilon$, which is included in the resultant AFT model as a parameter.
In the language of accelerated failure time models, this can be expressed in terms of the accelerated failure time assumption~\cite{kleinbaum1996survival}
\begin{equation} \label{assumption}
    S_\theta(t) = S_0 \left( \frac{t}{\phi_\theta(x)} \right),
\end{equation}
where $\phi_\theta$ is the acceleration factor, described by the joint effect of the covariates, such that
\[
    \phi_{\theta}(x) = \exp{ \left( \theta_{0} x_{0} + \theta_{1} x_{1} + \cdots + \theta_{n} x_{n} \right) },
\]
where $x=(x_0, \ldots, x_n)$ is a vector of covariates, $\theta=(\theta_0, \ldots, \theta_n)$ describe the fitted parameters. 
That is, as attack strength increases the time-to-failure decreases.

Any attack that produces a measurable degradation in model performance over time can be modelled using an AFT formulation, assuming there is a multiplicative relationship between explanatory covariates (\textit{e.g.,} attack strength) and time-to-failure~\cite{meyers_aft,aft_models,kleinbaum1996survival}.

\subsection{Choosing the best AFT Model}
\label{best-fit}
To choose a best-fit from a number of possible AFT functions, one should prepare the collected metrics and scores and then compare them using, \textit{e.g.,} Akaike Information Criterion (AIC), Bayesian Information Criterion (BIC), or Concordance, as per the best practices for this methodology~\cite{aft_models,kleinbaum1996survival}.
For AIC and BIC, that means choosing the model giving the smallest value. 
Concordance, however, is a number between $0$ and $1$ that quantifies the degree to which the survival time is explained by the model, where a $1$ reflects a perfect explanation~\cite{kleinbaum1996survival}, 0.5 reflects random chance, and 0 reflects a model that is perfectly wrong (a sign error). 
By evaluating $\mathbb{E}_{S_\theta}[T]$ under extreme perturbations, one can test the model and minimise the number of evaluated samples~\cite{aft_models,kleinbaum1996survival} rather than relying on $> 10^{12}$ validation samples, as required by IEC61508~\cite{IEC61508}.
The integrated calibration index (ICI) as well as the error at the 50th percentile E50~\cite{ici} are then calculated. 
These are the mean absolute difference between observed and predicted probabilities and the median absolute difference between observed and predicted probabilities, respectively~\cite{ici}.
Additionally, the data were split into a training set that was used to fit the model (80\%) and an unseen validation set (20\%), and the concordance, ICI, and E50 were measured for both.

\section{Cost and Robustness Metrics for Hardware Adaptation}
\label{cost}

Building on the survival analysis in Section~\ref{aft}, this section defines the cost and robustness metrics that inform the model hyper-parameter selection process. 
Cost is considered across three dimensions: computational efficiency (TRASH score~\cite{meyers_aft}), monetary deployment cost, and energy consumption. 
Together these metrics allow us to quantify the cost of training, deployment, and attacking a given model~\cite{IEC61508} on a given hardware configuration: the benign and adversarial accuracies (see Equations~\ref{eq:adv_success}~and~\ref{eq:acc}), the model training time, $t_{t}$, the model inference time or latency, $t_{i}$, the attack generation time, $t_{a}$, the cost per hour for a particular hardware, $C$, as well as the power consumption, $P$, of each tested model and attack.

\subsection{Accuracy}

Under the TRASHFIRE framework~\cite{meyers_aft}, adversarial accuracy serves as a measure of survival time across a specified time period, as outlined in Section~\ref{survival_time}.

\subsection{Training Time}

The training time, $T_t$, is the time it takes to evaluate $n$ samples, where $t_t$ is the training time per sample. It is defined as
\[
    T_t := t_t \cdot n  \cdot m,
\]
where $m$ is the number of epochs.

\subsection{Latency}

Latency is the time it takes to respond to a query. We assume that latency per batch is
\[
    T_i := t_i \cdot n,
\]
which will be driven by the memory bandwidth (measured in bits/second) of a given CPU or GPU and the size~\cite{vgg} and complexity~\cite{resnet} of a given neural network architecture.

\subsection{Attack Generation Time}

We distinguish between (Equation~\ref{eq:attack_time}) the computational cost of generating (both failed and successful) attacks and 
(Equation~\ref{eq:aft_time}) the stochastic time until a model failure occurs under attack.

Assuming a constant attack generation time per sample, the empirical average 
attack time over $n$ samples is
\begin{equation}
    t_a := \frac{T_a}{n},
    \label{eq:attack_time}
\end{equation}
where $T_a$ denotes the total time required to generate $n$ adversarial samples.

In contrast, the time until a successful attack induces model failure is a random variable. 
Prior work~\cite{meyers_aft} models this quantity using 
survival analysis. 
Treating model and attack parameters as covariates $\theta$, 
the expected time-to-failure can be expressed using an accelerated failure time 
(AFT) model such that
\begin{equation}
    t_a' \approx  \mathbb{E}_{S_\theta}[T] = \int_{0}^{T_a} S_{\theta}(t)\, dt.
    \label{eq:aft_time}
\end{equation}

We denote this expected survival time by $t_a'$, emphasizing that it captures 
the expected time until adversarial failure. 
Accordingly, $t_a$ and $t_a'$ 
represent fundamentally different quantities and are not directly comparable 
without additional assumptions linking attack generation and success processes.

\subsection{TRASH Score}
With an estimate of the expected survival time in hand, we quantify the cost-normalised failure rate, or the ratio of training time to attack time. 
Assuming that the cost scales linearly, as discussed above, the model builder's cost is proportional to the training time ($C_t \propto t_{t}$), and the attacker's cost is proportional to the attack time, which is approximately the expected survival time ($C_{a} \propto t_{a}' \approx \mathbb{E}_{S_\theta}[T]$). 
Using the definition of $\varepsilon$ from Equation~\ref{eq:adv_success} and definition of $t_a'$ from Equation~\ref{eq:aft_time}, we can express the cost of failure in adversarial terms as a function of per-sample inference time ($t_t$) and per-sample attack time ($t_a$)~\cite{meyers_aft},
\begin{equation}
	\textrm{TRASH}\approx \frac{t_t}{\mathbb{E}_{S_\theta}[T]} = \frac{t_t}{t_a'},
	\label{eq:cost}
\end{equation}
where a value larger than one indicates a model that is cheaper to break than it is to train.

\subsection{Monetary Deployment Cost}

Furthermore, we approach the cost of deployment at two scales. 
Firstly, we consider the cloud-rental scale, where a small business might test and deploy a model using, \textit{e.g.,} the Google Cloud Platform (GCP) compute costs as a measure of total cost. 
However, at a certain scale or with certain applications, it is more appropriate to talk about cost in terms of power (\textit{e.g.,} to deploy a self-driving car with a useful operating range). 

Finally, we define metrics that provide an efficient way to minimise the latency and cost of deployment, and to maximise the generalised performance of a model. 
We define the training cost as
\[
    C_t = C_h \cdot T_t,
\]
where $C_h$ is the cost per unit time for the hardware, $T_t$ is the training time.
Inference and attack costs follow analogously, denoted with $i$ and $a$ subscripts respectively.

\subsection{Power}

The power consumption for a particular piece of hardware, $P_h$, measured in Watts (Joules per second), can be thought of similarly such that the total power consumption of model training, inference, and attack is denoted with $t$, $i$, and $a$ subscripts respectively. 
In this work, power was monitored in real time using KEPLER~\cite{amaral2023kepler}, providing the energy cost metrics that inform the MAPE-K loop.

\section{Experiments}
\label{experiments}

This section details the implementation of the methodology in Sections~\ref{aft}--\ref{cost} to estimate the cost, power consumption, and adversarial robustness of ML models across the hyper-parameter and hardware search space, enabling rapid rejection of poor configurations and informed planning for self-adaptive cloud-native systems.

\subsection{Cloud Platform and Hardware}
To conduct the experiments and have access to different types of hardware, we utilised GCP\@.
Six virtual machines running Container-Optimised OS provided by GCP constituted the test-bed.
Using Google Kubernetes Engine 1.27.3 and Containerd 1.7.0, a cluster consisting of six worker nodes was created.
Three worker nodes were responsible for running the monitoring platforms --- Prometheus 2.47.2 and Grafana 10.2.0.
These nodes were of the ``e2-medium'' instance type provided by GCP\@.
In total, three GPU architectures were used --- the Nvidia P100, V100, and L4\@.
For P100 and V100 GPUs, the ``n1-standard-2'' type was used for the nodes and for L4 GPUs the ``g2-standard-4'' was used.

To assess the energy consumption of the experiments, KEPLER was deployed on each node to measure the power consumption of each experiment~\cite{amaral2023kepler}.
KEPLER collects per-pod energy metrics within the Kubernetes cluster and exports them as Prometheus metrics~\cite{sedghpour@ebpf}. 
Finally, all of the experiments were restricted to a \$1,000 budget (a research grant from Google). 
Approximately 10\% was used for development and 90\% for the evaluations.
Over the course of the evaluations, 6\% of the budget went to storage, 6\% to monitoring, and the remaining 88\% went to the GPUs.

\subsection{AFT Models}

For each hardware configuration and dataset (see Section~\ref{datasets}), the Tree-structured Parzen Estimator (TPE) algorithm was used to select the parameters~\cite{ozaki2020multiobjective,zitzler2008quality}, and it was iterated for 1,000 trials.
We used three optimisation criteria: benign accuracy, training time, and adversarial accuracy, seeking to maximise both benign and adversarial accuracy while minimising training time (and therefore deployment cost). 
We selected a set of parameters as per the TPE algorithm and trained on 80\% of the samples for each of the MNIST, CIFAR10, and CIFAR100 datasets. 
Of the remaining samples, 100 were withheld to be attacked and used to evaluate the adversarial accuracy. 
For each dataset, we tested this on 10 random splits of the data to create 10 unique train/validation pairs. 
For each trial, we recorded attack generation time, model training time, model inference time, benign accuracy and adversarial accuracy, and the size of the training set, validation set, and attack set. 
Using these values, we fit an AFT model to the number of failures (indicated by accuracy and sample size) and the attack generation time after adding categorical variables for the dataset and hardware device.

\subsection{Datasets}
\label{datasets}
The AFT models were evaluated using the MNIST~\cite{mnist}, CIFAR10~\cite{cifar}, and CIFAR100~\cite{cifar} datasets, chosen primarily for their standardised use in adversarial analysis~\cite{madry2017towards,croce_reliable_2020,carlini_towards_2017,deepfool} and decades of experimental results.
Before training, we centred and scaled the data so that the attack distance would be comparable for all tested datasets. 
Furthermore, to reduce the complexities of system overhead, distributed or federated training, and the effect of shared cloud environments, we restricted ourselves to datasets that were small enough to reside entirely within GPU memory with the model--- testing these more complicated configurations is left to future work though the general methodology would remain unchanged.

\subsection{Models}

The evaluations were restricted to a single model. 
Primarily, this was done to meet the budgetary constraints since evaluating more models would mean evaluating fewer pieces of hardware. 
As exact verification of adversarial robustness is computationally intractable---being NP-hard even for common classes of neural networks~\cite{carlini_towards_2017,ruan2019global}--- therefore, we tested each model and attack over a large hyper-parameter space.
So, we sampled learning rates in $[10^{-6}, 1]$, batch sizes in $[1, 10^5]$, and epochs in $[1, 50]$ for MNIST and CIFAR10 on the P100 and V100. For CIFAR100, the range of the tested epochs was increased to be in $[1, 100]$. 
The Feature Squeezing defence~\cite{feature_squeezing} was used to evaluate the efficacy of different bit-depths on the L4 hardware, as provided by IBM's adversarial robustness toolbox~\cite{art2018} with bit depths in $[4,8,16,32,64]$ which casts the inputs into \texttt{pytorch}-compatible arrays. 
Feature Squeezing was applied only to the L4 because that GPU's tensor cores natively execute 8-bit operations; applying the same bit-depth reduction to the P100 or V100 would instead emulate 8-bit arithmetic in software on 32-bit silicon, which does not reflect those GPUs' actual operating mode.
Model parameters were chosen using the \texttt{optuna} optimisation framework, the configuration was handled by \texttt{hydra}, and \texttt{dvc} was used to ensure reproducibility and aid in collaborative development.

\subsection{Attacks}

To examine the effect of model parameters at run-time, evasion attacks, which attack the model at the prediction stage, were examined.

The goal of using these attacks is to model the effect of noise on model performance rather than imagine a hypothetical 
Prior research~\cite{meyers,meyers_aft} has shown that the fast gradient method (see Eq.~\ref{eq:fgm}) is consistently the most effective at inducing a large number of failures in a small amount of time.
To evaluate the effect of adversarial noise (or lack thereof) on the samples, the noise levels were varied $0 < \|\varepsilon_i\| \leq 1$, sampled randomly from a uniform distribution.
This was done using the \texttt{adversarial-robustness-toolbox} package maintained by a team at IBM~\cite{art2018}.

\subsection{GPU Configurations}

Several hardware configurations were tested that had various hourly costs, peak power demands, and theoretical memory bandwidths. 
The V100 was chosen as a baseline, since it is routinely used in the literature~\cite{svedin2021benchmarking,xu2018deep}. 
The P100 architecture comes from the same line of server-grade GPUs, but from an older generation. 
The L4, however, is advertised as a machine built for inference, not training, relying on a smaller number of bits per tensor core. Consequently, the number of operations per second depends on the bit depth of the data and model weights, with peak numbers outlined in Table~\ref{tab:hardware} for 8-bit inputs. 
The rental cost of the hardware, measured in United States Dollars per hour, indicates the operating cost of a given model. 
To calculate this, the price per hour from each cloud service pricing page was used~\cite{pricing_1,pricing_2}, and the cost of training ($C_{t}$) and the cost of inference ($C_{i}$) were calculated from the cost of hardware ($C_{h}$), the training time ($T_{t}$), and the inference time ($T_{i}$).
\begin{table*}[h]
    \centering
    \begin{tabular}{lrrr}
    \toprule
                            & V100   & P100   & L4    \\
    \midrule
    Cost (USD/hour)         & 2.55   & 1.60   & 0.81  \\
    Power (Watts)           & 250    & 250    & 72    \\
    Memory Bandwidth (GB/s) & 900    & 732    & 300   \\
    % \bottomrule
    \end{tabular}
    \caption{Hardware specifications for the tested GPUs. The specifications were retrieved from Nvidia's website at the following links:
    \href{https://images.nvidia.com/content/technologies/volta/pdf/volta-v100-datasheet-update-us-1165301-r5.pdf}{V100 Datasheet},
    \href{https://images.nvidia.com/content/tesla/pdf/nvidia-tesla-p100-PCIe-datasheet.pdf}{P100 Datasheet}, and
    \href{https://nvdam.widen.net/s/rvq98gbwsw/l4-datasheet-2595652}{L4 Datasheet}. Prices were retrieved from \href{https://cloud.google.com/pricing/list}{Google Cloud Platform} for the \texttt{europe-west4} region on 3 December 2023.
    }
    \label{tab:hardware}
\end{table*}

\subsection{Survival Analysis}

In addition to the optimisation criteria of benign/adversarial accuracy and training time, prediction times, attack times, power consumption, batch size, attack noise, and number of epochs were also collected to be used as covariates in the AFT model, $S_{\theta}(t)$. 
To fit the AFT model and to plot the effect of the covariates, the \texttt{lifelines} Python package was used~\cite{lifelines}. 
The Weibull, Log Logistic, and Log Normal AFT models (see Section~\ref{aft}) were compared as per Section~\ref{best-fit}.

\section{Results and Discussion}
\label{results}

The experiments demonstrate that the proposed methodology provides reliable and cost-effective estimates of adversarial robustness across diverse hardware configurations. 
\subsection{Accuracy}
\label{res:acc}

Figure~\ref{fig:acc} shows the benign (left) and adversarial (right) accuracies for all datasets and hardware. 
It demonstrates little to no change in accuracy or adversarial accuracy, regardless of hardware. 
The benign accuracy decreases with difficulty (CIFAR10 \textit{vs.}~MNIST) or with the number of classes (CIFAR10 \textit{vs.}~CIFAR100). 
For all three datasets, the adversarial accuracy becomes the reciprocal of the number of classes (\textit{i.e.}, the accuracy we would expect with random data), demonstrating the efficacy of the attack outlined in Section~\ref{attacks}. 

\subsection{Time, Power, and Monetary Cost}
\label{res:cost}
\begin{figure*}[h]
    \centering
    \includegraphics[width=.8\textwidth]{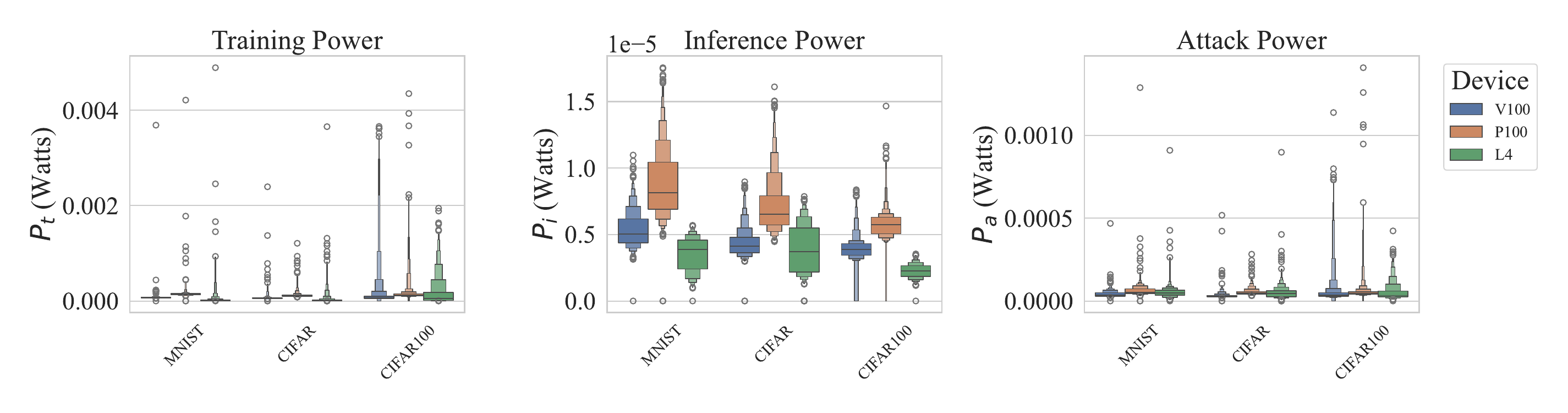}
    \caption{Here we depict the power consumption during training, inference, and attack generation for all hardware and datasets for all 1,000 trials using plots that depict the distribution of the second axis values using the width of the plot. 
    The power per sample was assumed to be uniform across the batch of samples for each training, inference, or attack measurement.
    Each colour is a different device and the datasets are displayed along the first axis. 
    Outliers are denoted with a white dot. 
    For these plots, the second axes have been scaled by the number of samples for the sake of comparison.}
    \label{fig:power}
\end{figure*}

\begin{figure*}[h]
    \centering
    \includegraphics[width=.8\textwidth]{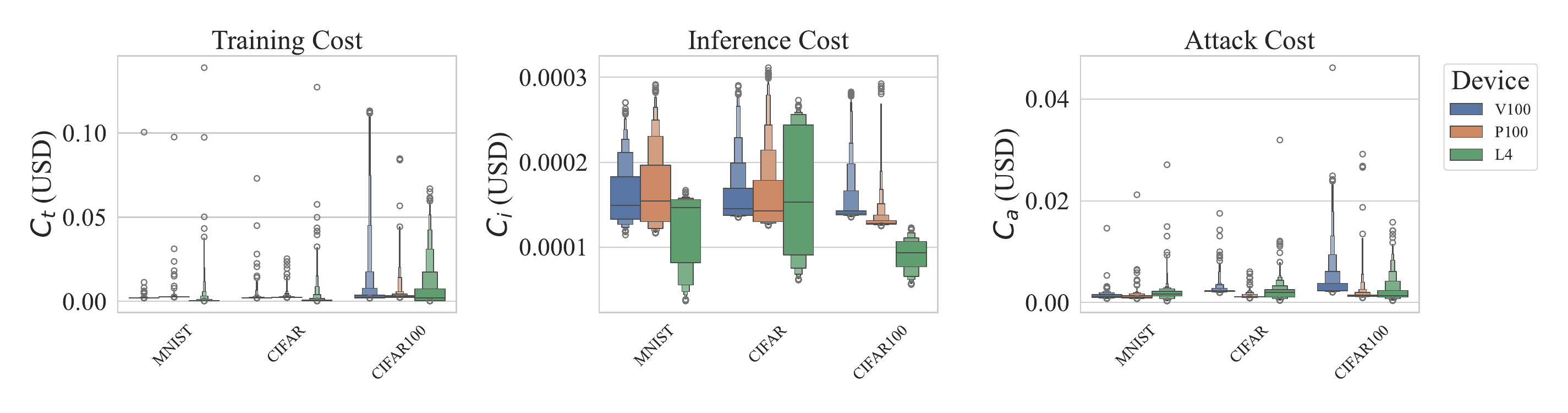}
    \caption{This depicts the monetary cost during training, inference, and attack generation for all hardware and datasets for all 1,000 trials using plots that depict the distribution of the second axis values using the width of the plot. 
    The cost per sample was assumed to be uniform across the batch of samples for each training, inference, or attack measurement.
    Each colour is a different device and the datasets are displayed along the first axis. 
    Outliers are denoted with a white dot. 
    For these plots, the second axes have been scaled by the number of samples for the sake of comparison.}
    \label{fig:cost}
\end{figure*}

The power consumption during training, inference, and attack generation for all hardware and datasets is illustrated in Figure~\ref{fig:power}. 
As expected, it tracks monetary cost (Figure~\ref{fig:cost}) closely --- since power is the predominant operating cost for data centres~\cite{dayarathna2015data}, cloud billing and power draw are structurally coupled by design. 

\begin{figure*}[tph!]
    \centering
    \includegraphics[width=.7\textwidth]{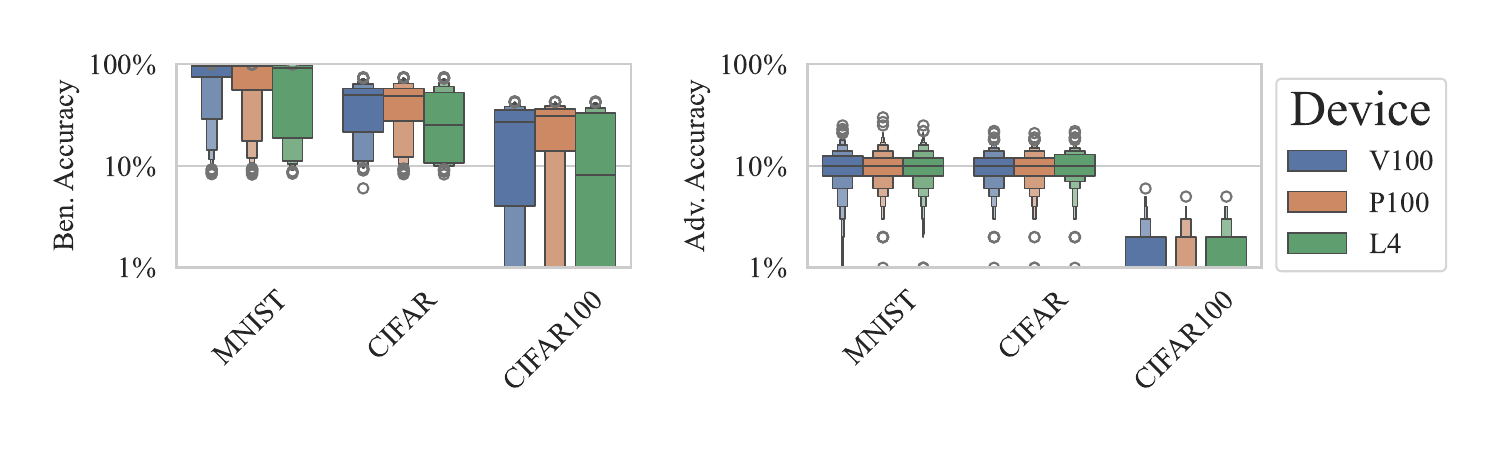}
    \caption{Benign and adversarial accuracy across all hardware and datasets for all 1,000 trials using plots that depict the distribution of the first axis values using the width of the plot. Each colour is a different device and the datasets are displayed along the first axis. Outliers are denoted with a white dot.}
    \label{fig:acc}
\end{figure*}

The monetary cost for each dataset and each piece of hardware is shown in Figure~\ref{fig:cost}. 
For all three datasets across all three pieces of hardware, the cost of training on a single sample often exceeds the cost of attacking a single sample. 
In the best-case scenario, they are comparable, but attacks consistently succeed with only 100 samples (Figure~\ref{fig:acc}) while model training requires orders of magnitude more.
Together, the power and cost measurements in Figures~\ref{fig:power}--\ref{fig:cost} constitute Monitor-phase observables that a MAPE-K Analyse phase can use to rank hardware configurations by cost-effectiveness.
%  However, we note that the attacker only needs to be lucky once, whereas the the model builder must be lucky always.

\begin{table*}[h]
\centering
\caption{
    Comparison of AFT Models across all hardware and datasets.
    }
\label{tab:aft_summary}
\begin{tabular}{lrrrrrrrr}
\toprule
             & AIC              & BIC & Conc & Val Conc & ICI & Val ICI & E50 & Val E50 \\
\midrule
Weibull      & $-2.11\cdot10^4$ & $-2.11\cdot10^4$ & 0.84 & 0.83 & 0.00 & 0.01 & 0.00 & 0.00 \\
Log Logistic & $-2.18\cdot10^4$ & $-2.18\cdot10^4$ & 0.84 & 0.83 & 0.01 & 0.01 & 0.00 & 0.00 \\
Log Normal   & $-2.25\cdot10^4$ & $-2.25\cdot10^4$ & 0.84 & 0.83 & 0.01 & 0.00 & 0.00 & 0.00 \\
\bottomrule
\end{tabular}
\end{table*}

\begin{figure*}[tph!]
    \centering
        \includegraphics[width=.28\textwidth]{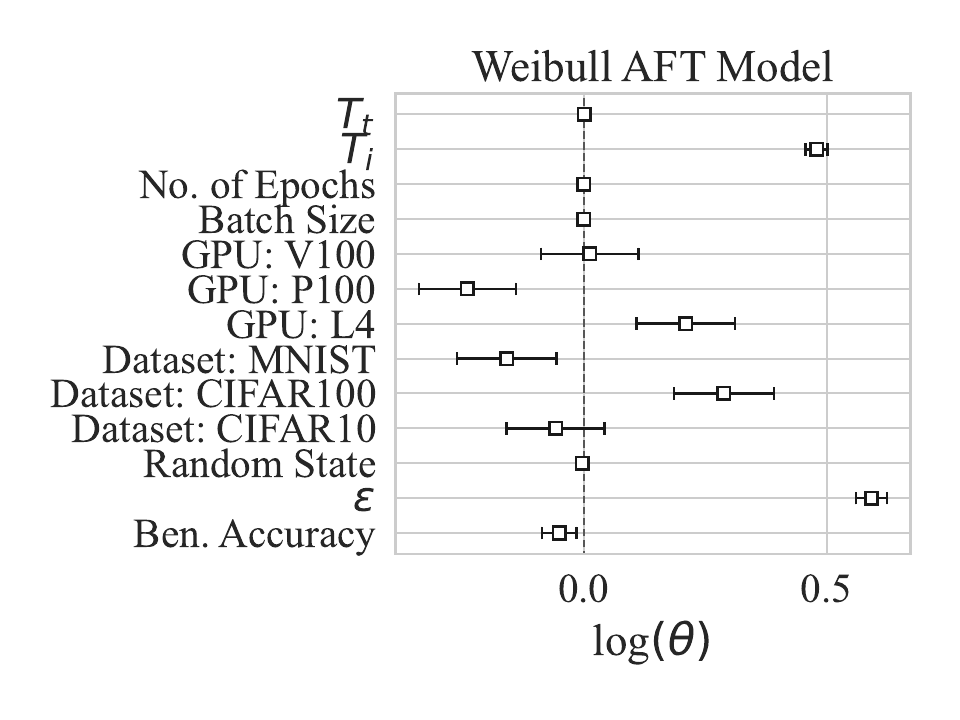}
        \includegraphics[width=.28\textwidth]{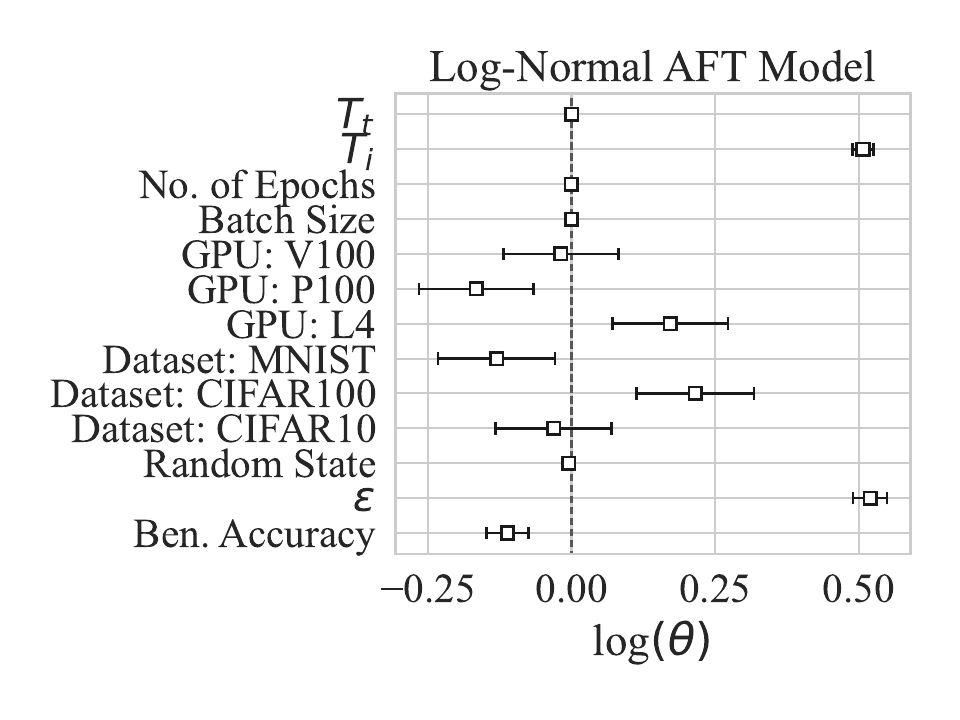}
        \includegraphics[width=.28\textwidth]{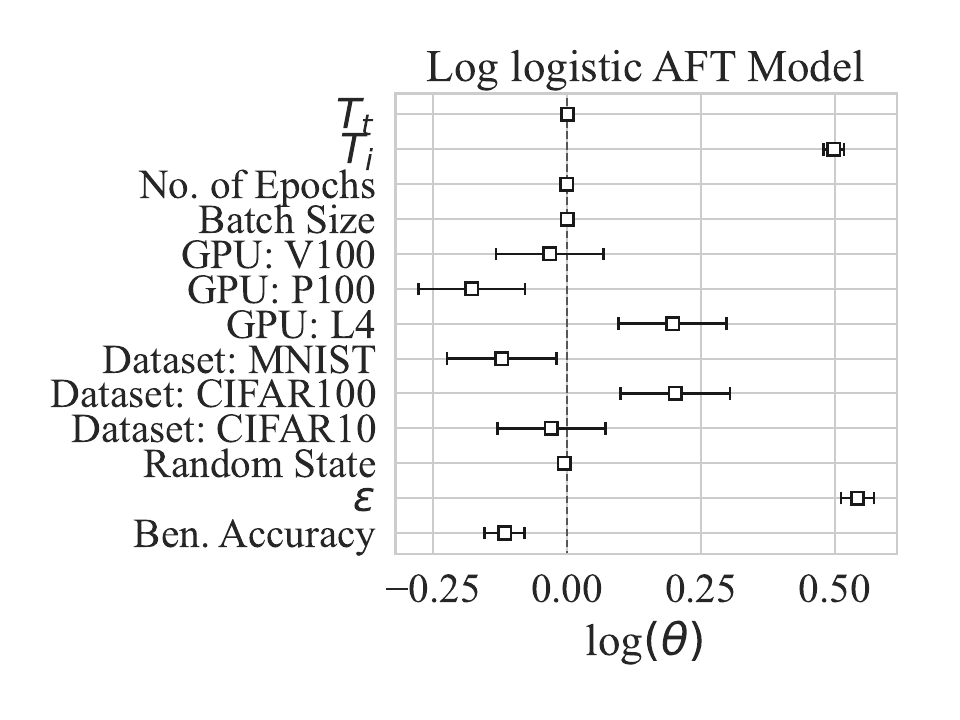}
    \caption{Coefficients of the Covariates for the Weibull, Log-Normal, and Log-Logistic AFT models. 
    Here, ``Random State'' is used as a control variable that should be (and is) close to 0. 
    A positive value indicates that a covariate increases the survival time and a negative value indicates that a covariate decreases the survival time. 
    The symbols $T_i$ and $T_t$ refer to training and inference time for all samples, while \textit{Ben. Accuracy} refers to the benign accuracy (accuracy on the un-altered samples), and $\varepsilon$ is the noise distance.}
    \label{fig:aft}
\end{figure*}
\subsection{AFT Models}
\label{res:aft}

Table~\ref{tab:aft_summary} shows the performance metrics for all three AFT models, as outlined in Section~\ref{survival_time}.
AIC~and~BIC are measures of goodness-of-fit, with a smaller value being preferred.
Concordance (Conc) is a value between 0 and 1 that reflects what proportion of events (failures) can be explained by the model.
ICI measures the average error between a cubic-spline and the model and E50 measures the median error between the spline and the model.
These are measured on both the train and validation sets of the measured data, with the latter being denoted ``Val''.
The concordance is strong ($>0.5$); similar for all three AFT models; and consistent across both the train and validation sets. 
We observed no more than 1\% mean absolute error in the probabilities ICI and no error in the median probability E50 across all three AFT models.
Figure~\ref{fig:aft} shows the log-scale coefficients for the AFT model. 
It shows that epochs, batch size, and training time have no effect on the survival time.
However, we can clearly see that inference time is as strong an indicator as the attack noise distance, revealing that model speed is nearly as important as the magnitude of the attack ($\varepsilon$).
Furthermore, we see that benign accuracy is negatively correlated with the survival time, confirming previous assertions that robustness ($S_{\theta}(t)$) is inversely related to benign accuracy (which is the standard indicator of generalisation performance)~\cite{carlini_towards_2017}.

Figure~\ref{fig:aft} further shows that hardware choice has a relatively small effect on robustness --- the L4 increases adversarial survival time by approximately 20\% and the P100 decreases it, both relative to the V100 --- despite the much larger disparity in cost outlined in Table~\ref{tab:hardware}.

\begin{figure*}[tph!]
    \centering
        \includegraphics[width=.31\linewidth]{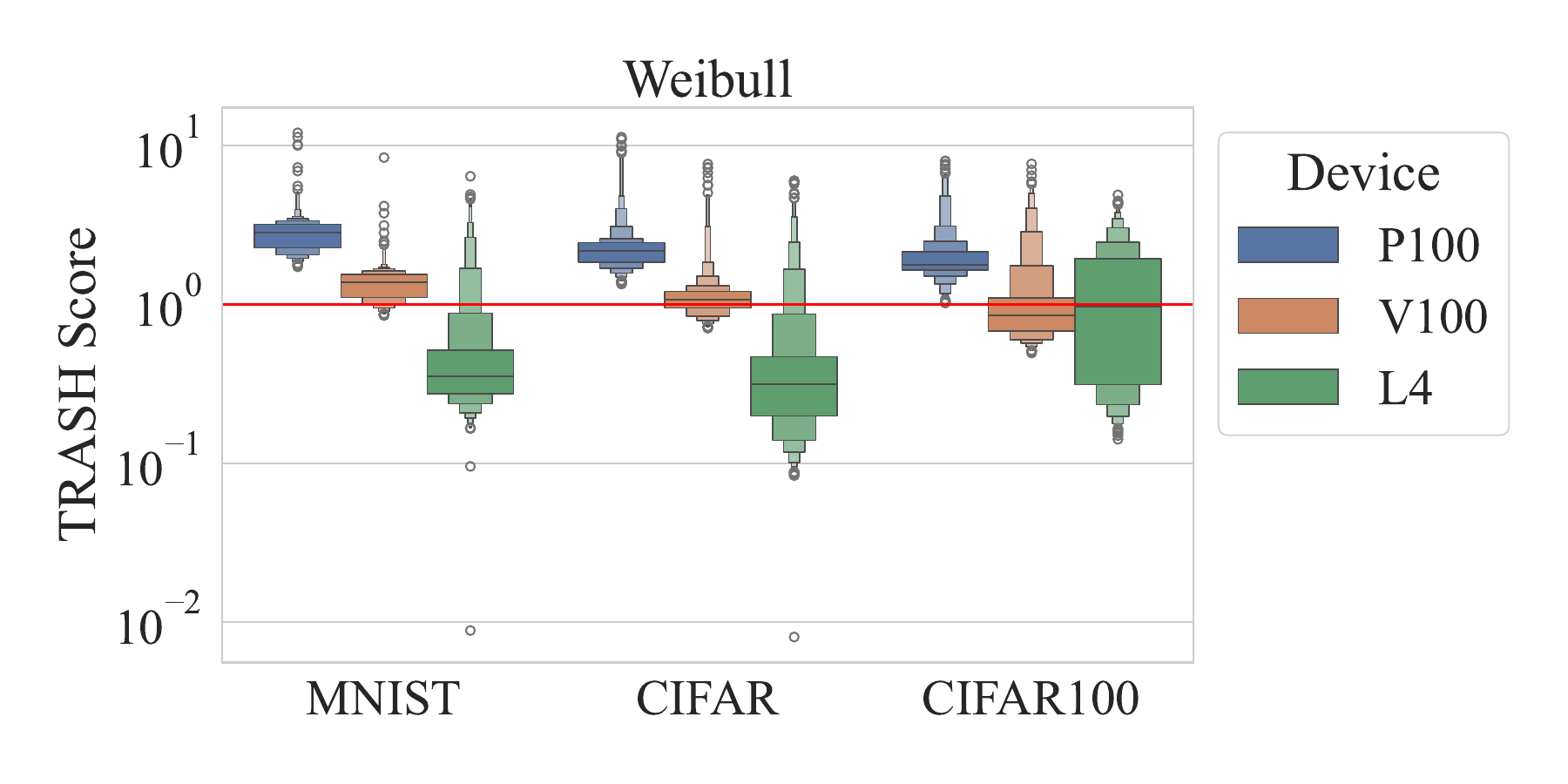}
        \includegraphics[width=.31\linewidth]{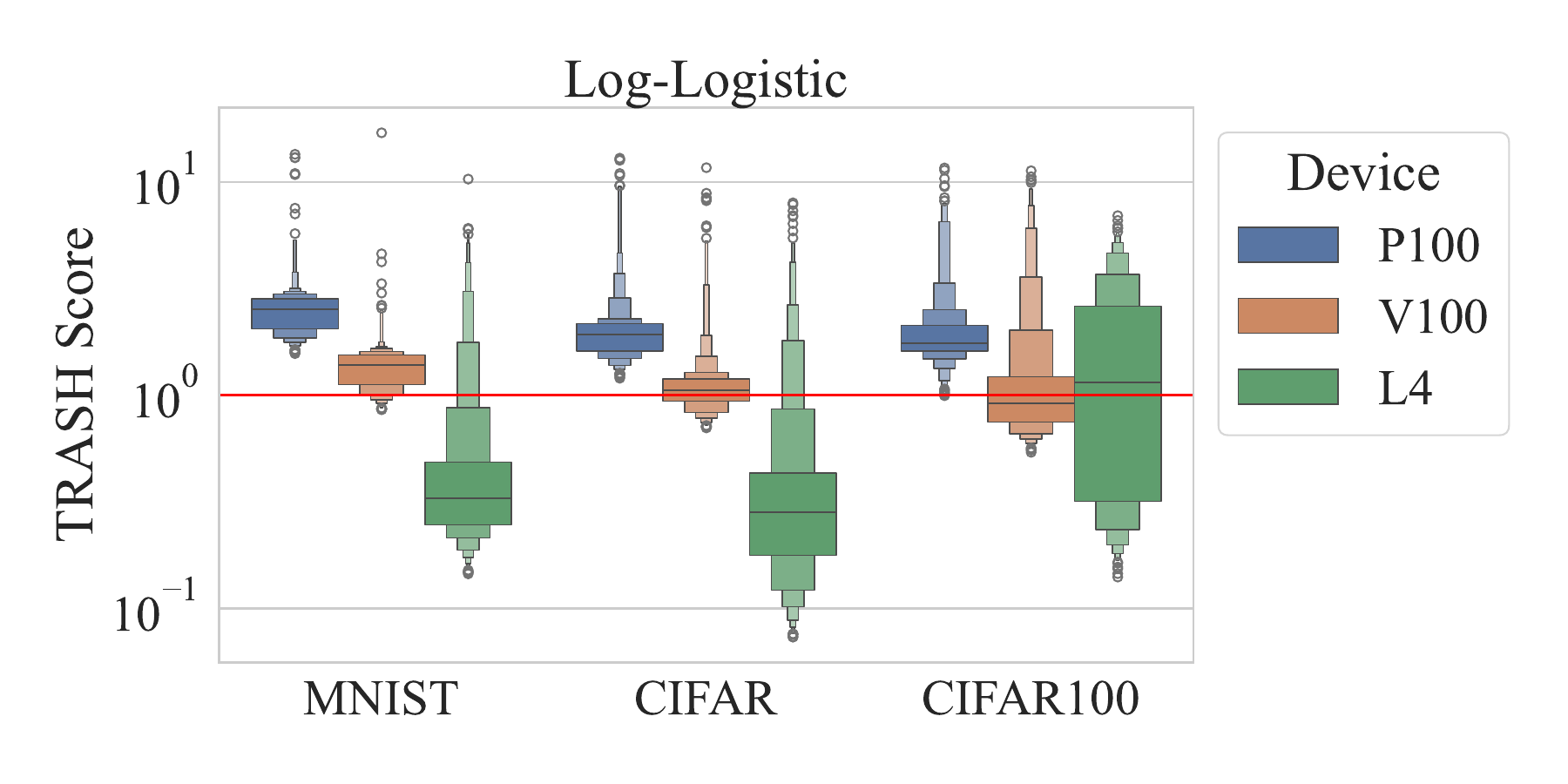}
        \includegraphics[width=.31\linewidth]{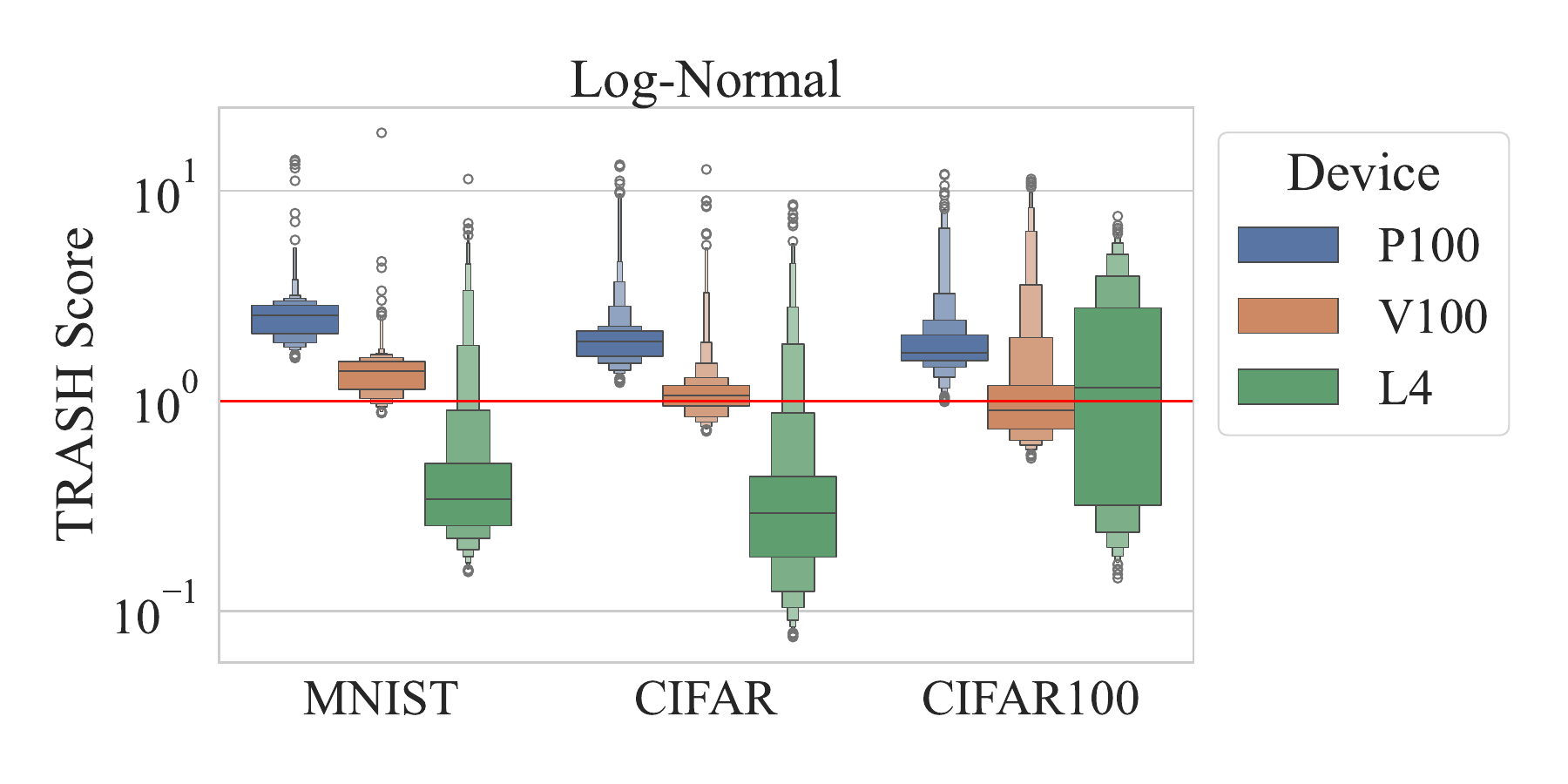}
    \caption{The \textit{training rate and survival heuristic} score (TRASH score) that depicts the per-sample ratio of training time to attack time for each of the tested AFT models.
    If this value is greater than one (the red line), then the model can be discarded as ineffective. 
    The first axis shows each dataset, the second axis shows the TRASH score and each GPU model is given its own colour. 
    Outlier scores are depicted with a white dot.}
    \label{fig:trash}
\end{figure*}

\subsection{Why Cost Matters}

Security analyses routinely assume optimistic scenarios for both attackers and defenders. 
In cryptography, these idealised attack and defence models are used to determine whether a scheme is computationally feasible to break~\cite{kamal2017study,leurent2020sha}. 
Likewise, if the cost of training a model greatly exceeds the cost of attacking it, then the model is effectively ``broken'' and should be discarded~\cite{meyers_aft}. 
Figure~\ref{fig:trash} shows the TRASH score (Eq.~\ref{eq:cost}), which measures the per-sample ratio of training to attack time.

Despite its lower memory bandwidth (Table~\ref{tab:hardware}), the L4 achieves superior cost (Table~\ref{tab:hardware}) and robustness (Figure~\ref{fig:aft}) because its tensor cores natively execute 8-bit operations, matching the 8-bit MNIST, CIFAR10, and CIFAR100 datasets. 
The additional precision of the P100 and V100 provides no useful information for these workloads. 
This result directly supports the Analyze phase of a MAPE-K loop by enabling self-adaptive systems to select the least expensive hardware configuration without sacrificing robustness and to discard models that are cheaper to attack than to train.

\subsection{Advantages of the Proposed Methodology}
\label{advantages}

Unlike conventional train/validation evaluation, whose precision is limited by the number of samples, the precision of survival-time estimates is determined by hardware clock resolution. 
Demonstrating the failure rates required by safety standards such as IEC~61508 (\textit{e.g.,} one failure in a million) would require millions of validation samples and repeated data collection after each software change~\cite{IEC61508}. 
In contrast, the proposed AFT model achieves an average error of approximately 1\% using only 10 sets of 100 samples (Table~\ref{tab:aft_summary}).

Because AFT models require relatively few samples, they can act as lightweight unit tests for ML components, estimating marginal risk per IEC~61508 without full-system integration or deployment~\cite{IEC61508,schmoor2000sample,lachin1981introduction}. 
Their strong predictive performance on unseen hyper-parameter configurations can also reduce the search space by eliminating candidates unlikely to improve survival time. Finally, the resulting cost and robustness estimates naturally support the planning stage of a MAPE-K control loop~\cite{kephart2003vision}, making the methodology suitable as a quantitative decision-support component for self-adaptive cloud-native ML systems.

\section{Scope and Limitations}
\label{considerations}

Although this work focuses on adversarial robustness, the proposed cost and survival analysis framework is general and can be applied whenever there is a multiplicative relationship between attack strength and model failure.

Timing measurements were designed to minimise jitter and account for GPU parallelism by assuming a uniform time-to-failure across samples within each benign or adversarial evaluation. 
While some samples or classes are likely easier to attack than others, we assume these effects average over the 100 attack samples. 
Although modelling class- or sample-specific failure rates is outside the scope of this work, it could reduce some of the remaining unexplained variance. 
Nevertheless, the strong agreement across AFT models (Table~\ref{tab:aft_summary}) suggests that the uniform-time assumption has little practical impact.

To minimise confounding factors, we restricted our experiments to models and datasets that fit entirely within GPU memory, isolating the effects of model configuration and hardware from secondary bottlenecks such as storage I/O, PCIe transfers, and distributed communication. 
While these factors can significantly affect latency and cost in production cloud deployments, they can be incorporated as additional covariates in the proposed survival analysis framework. 
The 24\,GB VRAM of the L4 is sufficient for standard 8-bit vision benchmarks (\textit{e.g.,} MNIST, CIFAR10, and CIFAR100); indeed, at the time of publication, 98.9\% of publicly available Kaggle datasets are smaller than 24\,GB~\cite{kagglehub}. 
Extending the methodology to distributed and federated settings remains future work.

The attack and training batch sizes were matched in every trial so that attack timing reflected typical deployment conditions. Since attacks operate on fewer samples than training, the additional parallelisation overhead likely causes attack efficacy to be slightly underestimated relative to benign measurements.

Finally, hyper-parameter optimisation was restricted to a single evasion attack (Eq.~\ref{eq:fgm}) due to computational constraints. 
However, the proposed AFT-based methodology naturally extends to other attack classes, including evasion, extraction, inversion, and poisoning attacks~\cite{carlini_towards_2017,biggio_poisoning_2013,choquette2021label,orekondy2019knockoff}.

\section{Conclusions}
\label{conclusion}

We presented an autonomic decision-support framework that applies survival analysis to cloud-native ML, enabling rapid, cost-efficient evaluation of hyper-parameter choices under adversarial noise. 
The results show that adversarial robustness is influenced far more by attack parameters than by reduced training times on newer or larger hardware.

The experiments demonstrate that the methodology is both sound and economical. 
Although GPU rentals accounted for 88\% of the cloud budget, replacing 32-bit data-centre GPUs (V100 and P100) with the 8-bit-optimised L4 reduced GPU costs by 75\% while improving both robustness and cost-effectiveness for standard image classification tasks. 
KEPLER proved similarly efficient, accounting for only 6\% of the total 
budget. AFT models achieved concordance scores above 0.83 on held-out data, confirming survival analysis as a reliable estimator of adversarial robustness. 
Attack cost consistently remained below training cost across all datasets and hardware, while the fitted coefficients identified inference latency ($T_i$)—rather than training time, model complexity, or epoch count—as the dominant predictor of robustness.

The coefficients also reinforce the well-known inverse relationship between benign accuracy and adversarial robustness. 
Even after accounting for GPU bandwidth (Table~\ref{tab:hardware}), the L4 remained the most robust and cost-effective platform (Figure~\ref{fig:trash}), despite being marketed primarily for inference.

Future work should extend the framework to black-box, poisoning, and membership inference attacks, validate its integration within live MAPE-K control loops, and evaluate distributed and federated deployments representative of modern large-scale ML systems.
% Prepared according to:
% https://www.overleaf.com/latex/templates/ieee-conference-template/grfzhhncsfqn
\bibliography{bibliography}{}
\bibliographystyle{plain}
\end{document}